\documentclass[journal]{IEEEtran}

\usepackage{amsmath}
\usepackage{graphicx}
\usepackage{hhline}
\usepackage[space]{cite}
\usepackage{notoccite}
\usepackage[thinlines]{easytable}
\usepackage{multirow}

\begin{document}

\title{A Review of Mixed-Effect Modeling in the Longitudinal Studies Using Medical Images of Patients}
\author{Fatemeh~Nasiri, Oscar~Acosta}
\maketitle

\begin{abstract}
In this review paper, some applications of the mixed effect modeling in medial image processing and longitudinal analysis is studied. For this purpose, a general structure is extracted from some of the researches in the literature. This structure includes a number of essential elements, each of which having a few design choices, namely 1) tracked features, 2) model's mathematical expression and random effects and finally 3) response prediction. Two research study examples in Alzheimer's disease and prostate tomography are also briefly introduced to further discuss the above design choices. 

\end{abstract}

% ========================================================
% ========================================================
% ========================================================

\section{Introduction}
\label{introduction}

\IEEEPARstart{W}{ith} the increasing prevalence of the signal processing applications in the medical science, the need for sophisticated automated processing tools is more perceived. One group of these applications deals with different imaging systems (e.g. MRI, CT, X-ray etc.), where one can derive useful information by studying big and diverse datasets of images taken from different patients in different stages of a disease. However, this diversity and plurality of data is helpful only if a proper and accurate automated tool is designed to process them.

Longitudinal study and disease tracking using medical imaging is one of the most important and effective processes during the medical treatment. Generally speaking, this procedure usually requires images regularly taken from patients' organ(s) involving in disease evolution. The information in these images hopefully guides to make better estimations regarding the future state of the organs and observe the impact of different drugs in different situations. For instance, one might be interested to discover the impact of a specific dosage of an anti-cancer drug on the size of the tumor through the time. Therefore, the question is how to benefit from the latent information in the temporally diverse set of medical images taken from different patients.

Statistical modeling and prediction is a practical and effective solution to address the above problem. The main idea is to exploit the repetitive behavior of the organ/tumors in response to a new condition (e.g. drug injection, aging etc.) and derive an accurate pattern. Like other similar machine learning and statistical modeling problems, this task requires a diverse image dataset with a reasonable size taken in different conditions. However, it has always been costly and more importantly dangerous to take excessive images from individual patients to specifically model their disease. The most feasible solution to this problem is to combine image datasets from different patients with the same disease condition and produce a rather general model. However, this simplification would compromise accuracy of the model due to the patient-specific behaviors of the treatments.

In this paper, we review the use of a statistical technique called mixed-effect modeling, to take into account patient-specific behaviors. As will be discussed, this technique helps the modeling process improve the accuracy in the longitudinal organ/tumor tracking. The rest of this paper is organized as follows. In section \ref{sec:framework}, the general framework of a prediction system is described in detail. In section \ref{sec:examples}, two example research projects with different prediction systems for tracking two different medical phenomenons have been studied. Finally, section \ref{sec:conclusion} concludes the paper with some discussions about the topic. 

% ========================================================
% ========================================================
% ========================================================
\section{General framework of a prediction system using mixed effect model}
\label{sec:framework}
In this section, we first explain principles of the mixed-effect modeling and its necessity in the modeling of a complex medical treatment process. Then, the main elements of a prediction system based on the mixed-effect modeling and their different choices are discussed.

\subsection{Principles of the mixed-effect modeling}
Basically, mixed-effect modeling or mixed modeling is a flexible statistical technique to exploit regularity in the pattern of a phenomenon which consists of both \textit{fixed} and \textit{random} effects. Herein, the term \textit{effect} refers to a system parameter that somehow impact the prediction value. This impact can be either non-random (i.e. fixed effects) or random (i.e. random effect) \cite{lindstrom1988newton} [cite mixed effect.pdf]. In matrix notation, a mixed-effect model can be represented as eq. \ref{eq:mixedmodelnotation}:

\begin{equation}
\label{eq:mixedmodelnotation}
y=f(\beta)+g(\gamma)+\varepsilon,
\end{equation}
where $f(\beta)$ and $g(\gamma)$ are fixed and random effects of the model and $\varepsilon$ is the prediction error.

\subsection{Key elements of a mixed effect modeling system}
By taking a look at the longitudinal studies with medical image processing, one can spot a few common essential elements. Here we briefly describe these elements \cite{ribba2014review}.
\subsubsection{Tracked features}
Depending on the disease under study, the tracked feature can be different. In the problems dealing with different types of cancers, the objective is to observe and model the growth in the size of tumor. ``Sum of longest diameters'' and ``Mean tumor diameter'' are two size-related features that can fairly represent the status of the tumor and also can be helpful in tracking it. According to the Response Evaluation Criteria in Solid Tumors (RECIST) \cite{tsuchida2001response}, these features are allowed to be measured and tracked for on a limited set of organs associated with lesions. 

However in other problems than the tomography, one might aim at monitoring more complicated features of the organ e.g. shape, displacement etc. For this purpose, some of the researches directly deal with pixels/voxels of the medical images taken from patients. In the simplest scenario, one might use each pixel/voxel of the image as one independent feature to track. More advanced approaches apply so-called feature reduction techniques such as Principle Component Analysis (PCA) to shrink the feature space \cite{rios2017population}.

There are also some other features that are used for tracking an organ's status through the time. For instance, in a research for studying the tumor growth rate, the Prostate-Specific Antigen (PSA) was tracked in order to monitor the prostate cancer status \cite{stein2010tumor}.

\subsubsection{Mixed-effect modelisation}
In the design of a mixed-effect model, there are different choices to make. One important choice is the mathematical expression used in modelisation. There are mainly two types of mixed-effect models in the literature: 
\begin{itemize}
\item Algebraic equations with the general form of:
\begin{equation}
y(t)=y_0+e^{-d.t}+g.t.
\end{equation} 
\item Differential equation with the general form of:
\begin{equation}
\frac{dy}{dt}=y_0+e^{-d.t}+g.t.
\end{equation}
\end{itemize}
In both equations, $y$ is the tracked feature, $t$ is time and $d$ is the model parameter. In this equation, the term exponential term of $e^{-d.t}$ is the random effect and the the term $g.t$ is the fixed effect in the model.

Once the mathematical expression is decided, one should decide about the number and the nature of random effects of the model. For example, here is a list of some popular random effects used in the literature of anti-cancer drug treatment:
\begin{itemize}
\item Drug-indudec decay of tumor
\item Net growth of tumor size
\item Tumor size nadir (the transition between decay and growth)
\item Drug concentration
\end{itemize}

\subsubsection{Response prediction}
As soon as the mixed effect model is trained with the training data, it can be used as a tool for prediction. This prediction generally includes estimation of future state of the tracked feature. The Expectation Maximization (EM) is one of the popular methods for addressing the response prediction problem.

% ========================================================
% ========================================================
% ========================================================
\section{Example researches using the mixed effect model based prediction framework}
\label{sec:examples}

After introducing the mixed effect modeling and its adoption in the longitudinal studies, here we briefly introduce two example researches in this domain. 
\subsection{A study on Alzheimer's disease by Opsina et al. \cite{ospina2012mixed}}
In this research, the authors have studied the temporal evolution the ``gray matter'', which is a very important indicator of progress in Alzheimer's disease. Gray matter is the darker tissue of the brain and spinal cord, consisting mainly of nerve cell bodies and branching dendrites \cite{sowell2001mapping}. The volume of this tissue is usually measured and tracked during treatment of Alzheimer's disease. 

The main purpose in this study is to model the impact of aging on the cortical gray matter quantification for two groups of subjects: healthy elderly (HE) and Alzheimer's disease (AD) patients. For this purpose, the high-level feature of gray matter volume is represented by low-level voxel-based morphology from MRI, PET and other biological markers. These voxels are then analyzed both in the temporal and spatial domains to build a mixed effect model. 

A non-parametric framework is used to model both groups of HD and AD subjects. For each normalized voxel $x$ from each subject $i$, where $i=\{1,...,n\}$, this framework is expressed as:
\begin{equation}
f_i(x,age)=m(x,age)+g(x,age).\Phi_i+e_i(x,age)+\varepsilon_i(x,age),
\end{equation}

where $m$, $g$ and $e_i$ are unknown functions and $\varepsilon_i$ is a random error. Moreover, the flag $\Phi$ is one if the $i$-th subject is an AD and is zero, otherwise. In the above equation, The term $m(x, age)+g (x, age)$ represents the expected density of the gray matter in an AD subject. Therefore, it can be understood that the role of flag $\Phi$ is to consider the gray matter emission in the AD group of subjects, due to their disease. Furthermore, $e_i (x, age)$ is deviation of the $i$-th subject from the voxel mean value corresponding to its group.

The analysis of this paper confirms that an age-dependent gray matter volume decrease usually happens both in HE and AD subjects. It also confirms that this decrease is more significant in the AD subjects than the HE subjects. Among the AD subjects, it was concluded that the rate of reduction in the gray matter become steeper with aging. 

\subsection{A study on Prostate cancer by Rios et al. \cite{rios2017population}}
In this research, the side-effects of radiotherapy in the prostate cancer treatment is analyzed using mixed effect modeling. The particular effect that has been studied in this paper is the bladder movement. The ultimate goal is to propose a mixed effect model, in order to enable an accurate and efficient bladder motion estimation under exposure of the cancer treatment drug. In contrast to the previous research example, here all the subjects are patients and the proposed approach tries to effectively model inter- and intra-patient behaviors of the bladder. Hence, the inter-patient features and the intra-patient features are considered as the fixed effects and random effects of the model, respectively.

From a high-level point of view and considering the framework introduced in section \ref{sec:framework}, the proposed model of this paper can be summarized in the following 5 steps.

\subsubsection{Imaging }
The very first step of the proposed process is the computed tomography imaging. In this step, raw data is taken from a diverse set of patient s through the treatment period. As this data naturally contains enormous outlier and noise, the next steps are performed to refine it and extract the most useful information from it.

\subsubsection{Spatial normalization of the input images  }
In this step, raw images taken in step one are processed to clip and extract the region of interest (i.e. the bladder). For this purpose, the three-dimensional space of voxels are processed and shrank to a reduced space of $135\times215\times55$ voxels containing the bladder voxels. 

\subsubsection{Surface representation}
Each bladder image extracted from step 2 is a solid volume. Therefore, all voxels in the three-dimensional space which occur inside that volume can be ignored during the process. This is due to the fact that each 3D volume can be perfectly represented only by using its outer surface. To this end, a mehtod called Spherical Harmonic (SPHARM) was adopted. SPHARM takes each image from the previous step in the $135\times215\times55$ space and projects it to a further reduced space of 272 voxels. 

\subsubsection{Feature reduction using PCA}
Considering the relatively small size of the dataset and also the fact that increasing it is extremely costly and harmful to the patients, one might further reduce the feature space. This step is necessary, since it is a proven rule in the machine learning that, when the size number of samples are limited, the size of feature space should be small enough to be trainable. 

The Principle Component Analysis (PCA) is a well-known method for efficient feature reduction. In this method, a statistical analysis is performed on the data and the target model, in order to detect uncorrelated part of the dataset. The expected output of this process is a reduced space which still contains a reasonably informative part of the data. In the research conducted in this paper, the PCA step reduces the feature space from 272 features to only 40 features. This set of 40 values, is considered as the final feature vector of each sample ans used in the rest of the mixed effect modeling process. 

\subsubsection{Mixed effect modeling }
According to the four previous steps, each image taken from patients is represented by a set of 40 features and the goal is to produce 40 independent mixed effect models associated to each feature. Let assume that there are $n$ patients in the dataset and number of $j_i$ ($i=1,...,n$) images have been taken from each patient. Also, let $z$ be the measure of motion/deformation that is aimed to be modeled (i.e. represented by the feature vectors). Therefore, it can be said that $z_{ijk}$ where $i=1,..,n$, $j=1,...,j_i$ and $k=1,...,40$ indicates the $k$-th feature of the $i$-th patient, taken at the $j$-th period of his/her treatment. The mixed effect modeling of this paper can be described as:
\begin{equation}
z_{ijk}=\mu_k+b_{ik}+\varepsilon_{ijk}.
\end{equation}

The $\mu_k$ term in the above equation represents the average of the $k$-th features among all patients in all of their images. The term $b_{ik}$ is a random variable representing deviation of the $i$-th patient from its mean $\mu_k$ and finally, $\varepsilon_{ijk}$ is a random variable representing deviation of the $i$-th variable from its personal mean. The these random variables follow the Gaussian distribution functions:

\begin{equation}
b_{ik} \sim N(0,\sigma_{bk}^2),
\end{equation}
and:

\begin{equation}
\varepsilon_{ijk} \sim N(0,\sigma_{k}^2).
\end{equation}

% ========================================================
% ========================================================
% ========================================================
\section{Conclusion}
\label{sec:conclusion}
In this paper, the use of mixed effect modeling in some of the medical image processing application was reviewed. To do this, first general philosophy behind the mixed effect modeling and its necessity in such applications is explained. Then the general principles and common elements of the different models in the literature was described. For this purpose, three different design elements in the process of mixed effect modeling were detected. These elements and their possible choices are summarized in Table \ref{tab:summary}.

% Please add the following required packages to your document preamble:
% \usepackage{multirow}
\begin{table}[]
\centering
\caption{Mixed effect model in the medical imaging applications: summary of different elements choices.}
\label{tab:summary}
\begin{tabular}{|l|l|}
\hline
Element                                                                             & Choices                                                                                                                                   \\ \hline
\multirow{2}{*}{\begin{tabular}[c]{@{}l@{}}Mathematical \\ expression\end{tabular}} & Algebraic equation                                                                                                                        \\ \cline{2-2} 
                                                                                    & Differential equation                                                                                                                     \\ \hline
\multirow{3}{*}{\begin{tabular}[c]{@{}l@{}}Tracked \\ features\end{tabular}}        & \begin{tabular}[c]{@{}l@{}}Size-based features, such as the \\ ``longest dumor diameter'' \\ or the ``mean tumor diameter''.\end{tabular} \\ \cline{2-2} 
                                                                                    & \begin{tabular}[c]{@{}l@{}}Voxel/Pixel-based features \\ following and optional \\ feature reduction step.\end{tabular}                   \\ \cline{2-2} 
                                                                                    & Other features line PSA.                                                                                                                  \\ \hline
\multirow{4}{*}{\begin{tabular}[c]{@{}l@{}}Random \\ effects\end{tabular}}          & Drug-induced decay in tumor.                                                                                                              \\ \cline{2-2} 
                                                                                    & Net growth of tumor size.                                                                                                                 \\ \cline{2-2} 
                                                                                    & Nadir tumor size                                                                                                                          \\ \cline{2-2} 
                                                                                    & Drug concentration                                                                                                                        \\ \hline
\end{tabular}
\end{table}

% ========================================================
% ========================================================
% ========================================================

\bibliographystyle{unsrt}
\bibliography{myBib}

\end{document}